\documentclass[letterpaper]{article} 
\usepackage{aaai23}  
\usepackage{times}  
\usepackage{helvet}  
\usepackage{courier}  
\usepackage[hyphens]{url}  
\usepackage{graphicx} 
\usepackage{natbib}  
\usepackage{caption} 
\usepackage{algorithm}
\usepackage{subcaption}

\usepackage{algorithmic}

\newcommand{\xhdr}[1]{\vspace{1.7mm}\noindent{{\bf #1.}}}

\usepackage{multirow}
\usepackage{booktabs, colortbl} 
\usepackage{amsmath}
\usepackage{amssymb}
\usepackage{makecell}
\usepackage{float}

\usepackage{hyphenat}
\usepackage{mathptmx}

\usepackage[english]{babel}
\usepackage{csquotes}
\usepackage{xcolor}
\usepackage{verbatim}
\usepackage{booktabs}
\usepackage{breqn}

\usepackage{tabularx}

\newcommand{\STAB}[1]{\begin{tabular}{@{}c@{}}#1\end{tabular}}

 \usepackage{booktabs}
\usepackage{caption}

\usepackage{newfloat}
\usepackage{listings}

\usepackage{xspace}
\usepackage{amsmath}
\usepackage{amsfonts}
\usepackage{url}
\usepackage{marvosym}
\usepackage{ifthen}

\newcommand{\chatoDisplayMode}[1]{#1}

\definecolor{MyRed}{rgb}{0.6,0.0,0.0} 
\definecolor{MyBlack}{rgb}{0.1,0.1,0.1} 
\newcommand{\inred}[1]{{\color{MyRed}\sf\textbf{\textsc{#1}}}}
\newcommand{\frameit}[2]{
  \begin{center}
  {\color{MyRed}
  \framebox[.9\columnwidth][l]{
    \begin{minipage}{.85\columnwidth}
    \inred{#1}: {\sf\color{MyBlack}#2}
    \end{minipage}
  }\\
  }
  \end{center}
}

\newcommand{\note}[2][]{\chatoDisplayMode{\def\@tmpsig{#1}\frameit{{\Pointinghand} Note}{#2\ifx \@tmpsig \@empty \else \mbox{ --\em #1}\fi}}}
\newcommand{\todo}[2][]{\chatoDisplayMode{\def\@tmpsig{#1}\frameit{{\Writinghand} To-do}{#2\ifx \@tmpsig \@empty \else \mbox{ --\em #1}\fi}}}

\newcommand{\abbrevStyle}[1]{#1}

\newcommand{\etal}{\abbrevStyle{et al.}\xspace}
\newcommand{\vs}{\abbrevStyle{vs.}\xspace}

\newcommand{\textcite}[1]{\citeauthor{#1} \shortcite{#1}}

\newcommand{\hide}[1]{}

\makeatletter
\newcommand{\iffont}[2]{\ifthenelse{\equal{\f@family}{#1}}{#2}{}}
\makeatother

\iffont{ptm}{
  \usepackage{mathptmx}

  \DeclareSymbolFont{greek}{OML}{cmm}{m}{n}
  \DeclareMathSymbol{\alpha}{\mathalpha}{greek}{"0B}
  \DeclareMathSymbol{\beta}{\mathalpha}{greek}{"0C}
  \DeclareMathSymbol{\gamma}{\mathalpha}{greek}{"0D}
  \DeclareMathSymbol{\delta}{\mathalpha}{greek}{"0E}
  \DeclareMathSymbol{\epsilon}{\mathalpha}{greek}{"0F}
  \DeclareMathSymbol{\zeta}{\mathalpha}{greek}{"10}
  \DeclareMathSymbol{\eta}{\mathalpha}{greek}{"11}
  \DeclareMathSymbol{\theta}{\mathalpha}{greek}{"12}
  \DeclareMathSymbol{\iota}{\mathalpha}{greek}{"13}
  \DeclareMathSymbol{\kappa}{\mathalpha}{greek}{"14}
  \DeclareMathSymbol{\lambda}{\mathalpha}{greek}{"15}
  \DeclareMathSymbol{\mu}{\mathalpha}{greek}{"16}
  \DeclareMathSymbol{\nu}{\mathalpha}{greek}{"17}
  \DeclareMathSymbol{\xi}{\mathalpha}{greek}{"18}
  \DeclareMathSymbol{\pi}{\mathalpha}{greek}{"19}
  \DeclareMathSymbol{\rho}{\mathalpha}{greek}{"1A}
  \DeclareMathSymbol{\sigma}{\mathalpha}{greek}{"1B}
  \DeclareMathSymbol{\tau}{\mathalpha}{greek}{"1C}
  \DeclareMathSymbol{\upsilon}{\mathalpha}{greek}{"1D}
  \DeclareMathSymbol{\phi}{\mathalpha}{greek}{"1E}
  \DeclareMathSymbol{\chi}{\mathalpha}{greek}{"1F}
  \DeclareMathSymbol{\psi}{\mathalpha}{greek}{"20}
  \DeclareMathSymbol{\omega}{\mathalpha}{greek}{"21}
  \DeclareMathSymbol{\varepsilon}{\mathalpha}{greek}{"22}
  \DeclareMathSymbol{\vartheta}{\mathalpha}{greek}{"23}
  \DeclareMathSymbol{\varpi}{\mathalpha}{greek}{"24}
  \DeclareMathSymbol{\varrho}{\mathalpha}{greek}{"25}
  \DeclareMathSymbol{\varsigma}{\mathalpha}{greek}{"26}
  \DeclareMathSymbol{\varphi}{\mathalpha}{greek}{"27}
  \DeclareSymbolFont{otone}{OT1}{cmr}{m}{n}
  \DeclareMathSymbol{\Gamma}{\mathalpha}{otone}{0}
  \DeclareMathSymbol{\Delta}{\mathalpha}{otone}{1}
  \DeclareMathSymbol{\Theta}{\mathalpha}{otone}{2}
  \DeclareMathSymbol{\Lambda}{\mathalpha}{otone}{3}
  \DeclareMathSymbol{\Xi}{\mathalpha}{otone}{4}
  \DeclareMathSymbol{\Pi}{\mathalpha}{otone}{5}
  \DeclareMathSymbol{\Sigma}{\mathalpha}{otone}{6}
  \DeclareMathSymbol{\Upsilon}{\mathalpha}{otone}{7}
  \DeclareMathSymbol{\Phi}{\mathalpha}{otone}{8}
  \DeclareMathSymbol{\Psi}{\mathalpha}{otone}{9}
  \DeclareMathSymbol{\Omega}{\mathalpha}{otone}{10}
  \DeclareSymbolFont{syms}{OML}{cmm}{m}{it}
  \DeclareMathSymbol{\partial}{\mathord}{syms}{"40}
  \DeclareMathAlphabet{\mathbold}{OML}{cmm}{b}{it}
  \DeclareSymbolFont{largesymbols}{OMX}{cmex}{m}{n}

}

\DeclareCaptionStyle{ruled}{labelfont=normalfont,labelsep=colon,strut=off} 
\floatstyle{ruled}
\newfloat{listing}{tb}{lst}{}
\floatname{listing}{Listing}

\usepackage{newfloat}
\usepackage{listings}
\DeclareCaptionStyle{ruled}{labelfont=normalfont,labelsep=colon,strut=off} 
\floatstyle{ruled}
\newfloat{listing}{tb}{lst}{}
\floatname{listing}{Listing}
\title{Reddit in the Time of COVID}
\author{
Veniamin Veselovsky, Ashton Anderson\\
}
\affiliations{EPFL, University of Toronto\\
veniamin.veselovsky@epfl.ch, ashton@cs.toronto.edu}

\usepackage{bibentry}

\begin{document}

\maketitle
\begin{abstract}
When the COVID-19 pandemic hit, much of life moved online. Platforms of all types reported surges of activity, and people remarked on the various important functions that online platforms suddenly fulfilled. However, researchers lack a rigorous understanding of the pandemic's impacts on social platforms, and whether they were temporary or long-lasting. We present a conceptual framework for studying the large-scale evolution of social platforms and apply it to the study of Reddit's history, with a particular focus on the COVID-19 pandemic. We study platform evolution through two key dimensions: \emph{structure} \vs \emph{content} and \emph{macro}- \vs \emph{micro}-level analysis. Structural signals help us quantify how much behavior changed, while content analysis clarifies exactly how it changed. Applying these at the macro-level illuminates platform-wide changes, while at the micro-level we study impacts on individual users. We illustrate the value of this approach by showing the extraordinary and ordinary changes Reddit went through during the pandemic. First, we show that typically when rapid growth occurs, it is driven by a few concentrated communities and within a narrow slice of language use. However, Reddit's growth throughout COVID-19 was spread across disparate communities and languages. Second, all groups were equally affected in their change of interest, but veteran users tended to invoke COVID-related language more than newer users. Third, the new wave of users that arrived following COVID-19 was fundamentally different from previous cohorts of new users in terms of interests, activity, and likelihood of staying active on the platform. These findings provide a more rigorous understanding of how an online platform changed during the global pandemic.\end{abstract}




\maketitle

\section{Introduction}


A major impact of the COVID-19 pandemic was a dramatic surge in online social platform usage. Virtually every major site reported large increases in activity, with many previously-offline facets of life necessarily being replaced by online analogs~\cite{feldmann2021year}. 
Although examples of this sea change are familiar from our own experience, we have yet to systematically quantify the structure and content of how social media changed during the course of the pandemic. 
This leaves significant gaps in our understanding of social media.
Did the processes governing online platforms before COVID-19 continue throughout the pandemic, or did a new organization emerge? Were the new users that joined post-2020 fundamentally different from previous cohorts of new users? What exactly were the changes that occurred, and did they last? 

Answering these questions is important for both theoretical and pragmatic reasons. 
From a theoretical perspective, rigorously measuring what happened on social media during COVID-19 is a prerequisite to understanding \emph{why} it happened. 
Although anecdotal experiences provide important early guidance in developing  understanding, only rigorous measurement of what occurred can provide a sufficiently solid empirical foundation for further theoretical inquiry. 
Furthermore, understanding how social media changed as a result of COVID-19 is important not only for understanding the effect of crises on mass communication but it is also a rare opportunity to understand the nature of social media more generally. 
Much like neuroscientists advance their understanding of the brain through analyses of unique traumatic brain injuries and their repercussions, we can better understand the ``hivemind'' of social media by measuring how it responded to the unique shock of a global pandemic. 
From a pragmatic perspective, quantifying how social media has (and hasn't) changed during this turbulent period will provide a baseline comparison for current and future analyses of social media activity in a post-pandemic world and help inform the design of tomorrow's digital environments.




Despite their importance, these questions have proven difficult to answer. 
In this work, we present a general approach for studying platform evolution that is applicable during either times of crisis or stability. Our approach (illustrated in Figure 1) defines two key dimensions of analysis: 1) structure versus content and 2) platform (macro-level) versus user (micro-level). 
Structural signals help us quantify how much behavior changed, while content analysis clarifies exactly how it changed. Applying these at the macro-level illuminates platform-wide changes, while at the micro-level we discover more nuanced user evolution. 
We apply this framework to the entire history of Reddit, a large social media platform used by millions of people both prior to and during the pandemic. 

\begin{figure}
    \centering
    \includegraphics[width=3.3in]{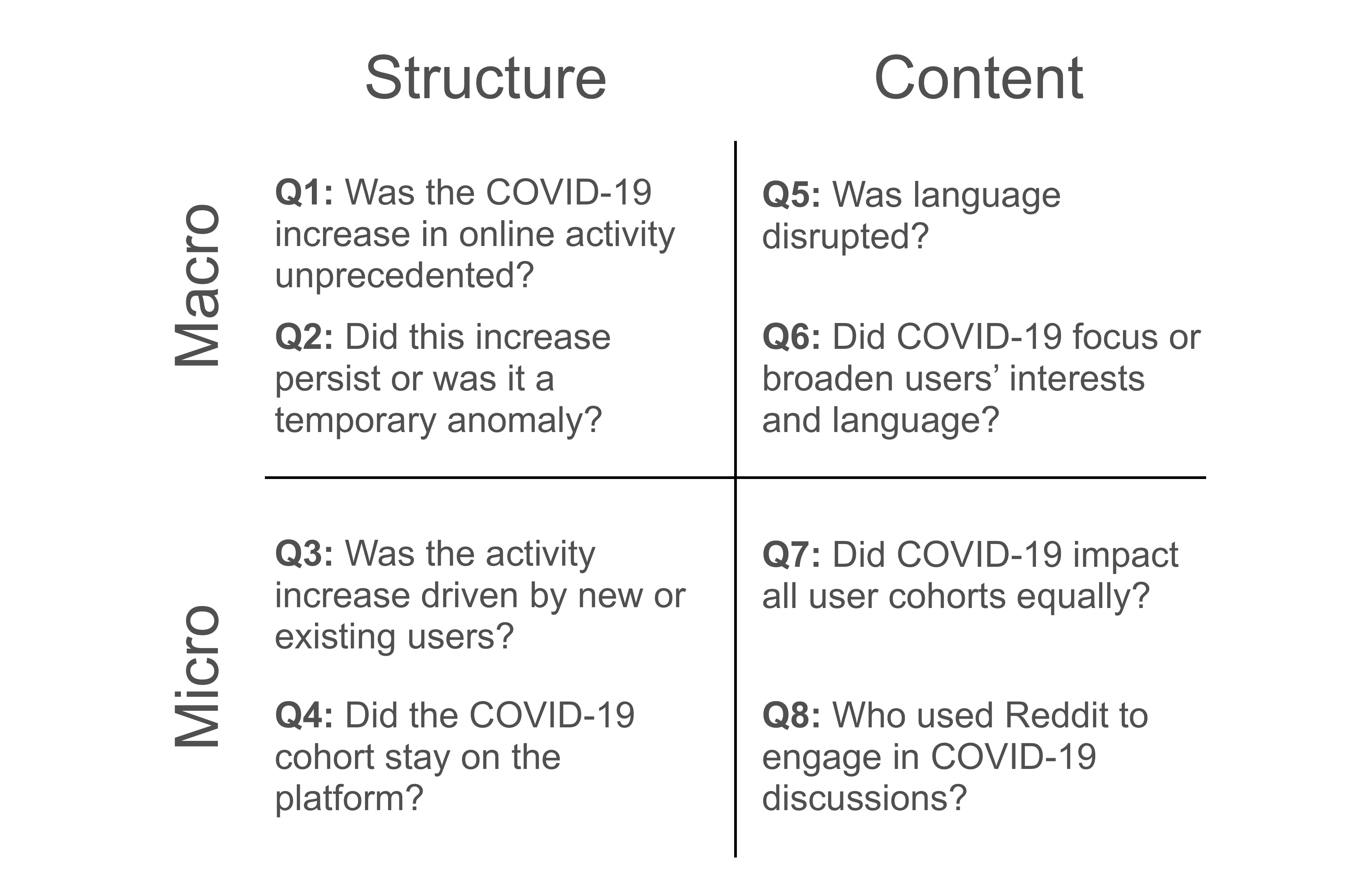}
    \caption{An illustration of our conceptual framework and our research questions.}
    \label{fig:questions}
\end{figure}






\xhdr{The Present Work} 
We showcase our general approach alongside the questions we answer in Figure \ref{fig:questions}. 
Our first focus of analysis is on the structure of Reddit at both the macro- and micro-levels. 
We find that Reddit experienced an immediate and significant increase in activity at the start of COVID-19. Overall activity jumped, new communities and topics of discussion emerged, and a large influx of users arrived, often disrupting the experience of previous cohorts. These new users that joined during the pandemic had different activity patterns and areas of interest; their early experiences differed from the traditional new user experience.    
In parallel, new language emerged, even beyond expected COVID-related vocabulary; new clusters of communities became popular; and the behaviors of older and newer users drifted. Government stimulus inspired investing discussions, quarantine increased political and world events discussions, and activity in NSFW communities doubled. These changes were sometimes short-lived, but others persisted. Measuring these dynamics is important for distinguishing between what aspects of online life have returned to our previous ``normal'' versus which aspects have become a ``new normal.''

Two years on, many of these changes have persisted. For example, the boost in monthly activity and the number of active subreddits did not wane and new users still differ from their pre-COVID counterparts. COVID-era language has also reached a natural background level on the platform, with words like ``coronavirus,'' ``stimulus,'' ``masks,'' and ``asymptomatic'' becoming commonplace. The most notable difference, however, lies in the types of communities that are popular. 

In summary, our main contributions are the following, and the rest of the paper discusses each in detail:
\begin{itemize}
    \item A conceptual approach to studying social media analyses separating structure from content and macro-level from micro-level analysis (Section \ref{data}).
    \item An analysis of the structural changes Reddit went through during COVID-19 (Section \ref{structure}).
    \item An analysis of the content-level shifts Reddit experienced by the platform and user-levels (Section \ref{content}).
\end{itemize}

\section{Related Work}
COVID-19 has affected our interactions with the digital world, augmenting everything from the number of bits that were communicated across the Internet to the infrastructures built on top of it. In this section, we touch on some of the existing literature that has captured some of these effects. 

\xhdr{Internet and social media}
Anja Feldmann et al. utilized data from a series of ISPs and IXPs to illustrate the mass influx in Internet use, which not only spiked with lockdowns, but exhibited a lasting and unequal growth depending on location and customer profile \cite{feldmann2021year}. 
A large swath of the COVID-19 social research has centered on the social media giants that acted as connectors and mediators for our at-home interactions. 
A cross-platform study was conducted tracking the number of users on various platforms, as well as the content they discussed, with a particular focus on potentially misleading link-sharing \cite{cinelli2020covid}. 
Other studies have similarly focused on various questions of misinformation and fake news, quantifying its size, detection, and user ``susceptibility''~\cite{evanega2020coronavirus,shang2022duo,khan2022detecting,teng2022characterizing,weinzierl2022identifying}.
Other work has instead adopted a reactive approach by studying how to combat misinformation through correcting user responses \cite{seo2022if} and examining how consumers should engage with COVID-related content \cite{raza2022fostering}. 

\xhdr{User changes} User-level studies have instead tracked how the individual experience has changed. This type of work typically examines how users or cohorts of users changed before and after the pandemic. Zhunis \etal found that shortly after the outbreak of COVID-19 users the amount of emotionally negative posts increased, but several months after, users returned to their pre-COVID emotion levels. In sum, users that were generally positive before the pandemic returned to being positive, and users that were generally negative returned to being negative~\shortcite{zhunis2022emotion}.  
A similar paper conducted a global analysis of sentiment prior and throughout the pandemic examining how lockdowns and COVID-19 cases affected people's well-being, where they utilized synthetic controls to study neighboring regions~\cite{wang2022global}. 
Other studies have instead focused on how the user composition has changed throughout the pandemic, like Wikipedia, which experienced a large influx in volunteer editing following COVID. \cite{ruprechter2021volunteer}.




\xhdr{Social media during times of crisis}
Social media and Web use in times of crisis has been historically an important area of study. Research has highlighted social media's role as a connector and a disseminator of information, giving people the power of sharing what occurs \cite{kogan2015think}. Another study examined Twitter's evolution across a few disasters by looking at the structural change and showing that users who joined during a crisis tend to be longer-lasting users~\cite{hughes2009twitter}. The question of what ``should'' social media's role be in times of crisis and how to educate people about it has been presented \cite{dailey2018crisis}.



\section{Data and Methodology}\label{data}
Our analysis focuses on Reddit, one of the largest social media sites which is composed of hundreds of thousands of communities (called subreddits) dedicated to  everything from politics (r/politics) to Japanese Woodworking (r/JapaneseWoodworking).  
This wide collection of communities, alongside Reddit's minimal algorithmic curation, makes it a good platform to study the various effects of COVID-19 across social media. 
Our data consists of all comments and submissions on Reddit from its inception in December 2005 to June 2022~\cite{baumgartner2020pushshift}.
This includes almost 13 billion comments by 89 million users and 5 billion submissions by 48 million users. 

Reddit's size comes with the potential for noise and spam. To account for this we took several pre-processing steps to limit bots and spam users. 
First, we defined a bot-account as any user with ``bot'' or ``mod'' (both case insensitive) in their username. 
Second, we de-duplicated comments by removing all comments with more than 100 characters that have the same text as more than two other comments (a variation of the approach adopted by Nobel \etal~\shortcite{noble2021semantic}). 
Finally, we also pre-processed the text of the comments for Section 5 by removing stop words, punctuation, numbers, and URLs, as well as dropping all comments with fewer than 3 characters.


\xhdr{Defining the start of the pandemic}
Since this paper compares Reddit before and during COVID-19, we need a point of reference to define the start of the pandemic. Past work has relied largely on case timelines or mobility restrictions data~\cite{ribeiro2021sudden,yang2022effects}. On Reddit, however, matching mobility restrictions with location data is infeasible since users do not publicly disclose their geographic location. 
Moreover, since Reddit is used all over the world, we cannot simply use the mobility restrictions from any one country. Even if we chose the United States, which is responsible for most Reddit traffic, we still would have state-by-state variations in mobility restrictions. For this reason, we rely on a simple platform-wide approximation for the start of the pandemic as of February 1, 2020. This errs on the side of caution by defining COVID-19 as being earlier than other estimates. Any changes we notice on Reddit are likely to only be exaggerated if we define the start of COVID-19 at a later date. 

\xhdr{Analysis framework} We propose an analysis framework for capturing platform changes over time using two axes: structure versus content (major axis) and macro- versus micro-level (minor axis). 
The structural analysis examines the meta-level factors associated with \emph{how} Reddit changes, including the size of the platform and the dynamics of its user base. 
On the other hand, the content analysis explores \emph{what} areas of attention users are focused on and the language they use.
Macro-level (platform-level) versus micro-level (user-level) is a complementary axis that instead focuses on changes occurring at different resolutions on the platform. 
Macro changes examine aggregated metrics across various dimensions, whereas micro-level changes capture the user-level mechanisms that may be driving some of the macro movements.

\xhdr{Effect estimation}
Comparing metrics before and during COVID-19 risks picking up on natural trends instead of COVID-related impacts. More generally, there exist natural changes that drive certain metrics on the platform which will confound the effects of COVID-19. 
To accommodate this possibility we will rely on a well-established time series prediction tool called Prophet, which captures non-linear trends, holiday effects, and ignores outliers~\cite{taylor2018forecasting}, and is often used in temporal analyses~\cite{zunic2020application,usher2020brexit}.
Our general setup is to fit Prophet for a metric on Reddit's historic pre-COVID data (up to February 2020) and then predict the future. 
Concretely, prior to February 2020, the predicted curve measures the prediction of month $T$, given months $t < T$. On February 2020, we stop the month-to-month prediction and estimate until June 2022. Then, we use the difference between real-world data (where COVID-19 happened), and the counterfactual forecast produced by Prophet to estimate the impact of COVID-19 on Reddit.
This setup helps us disentangle the natural growth in Reddit from that associated with COVID-19 and assess whether the change was long-lived.

\begin{figure}
    \centering
    \includegraphics[width = 3.3in]{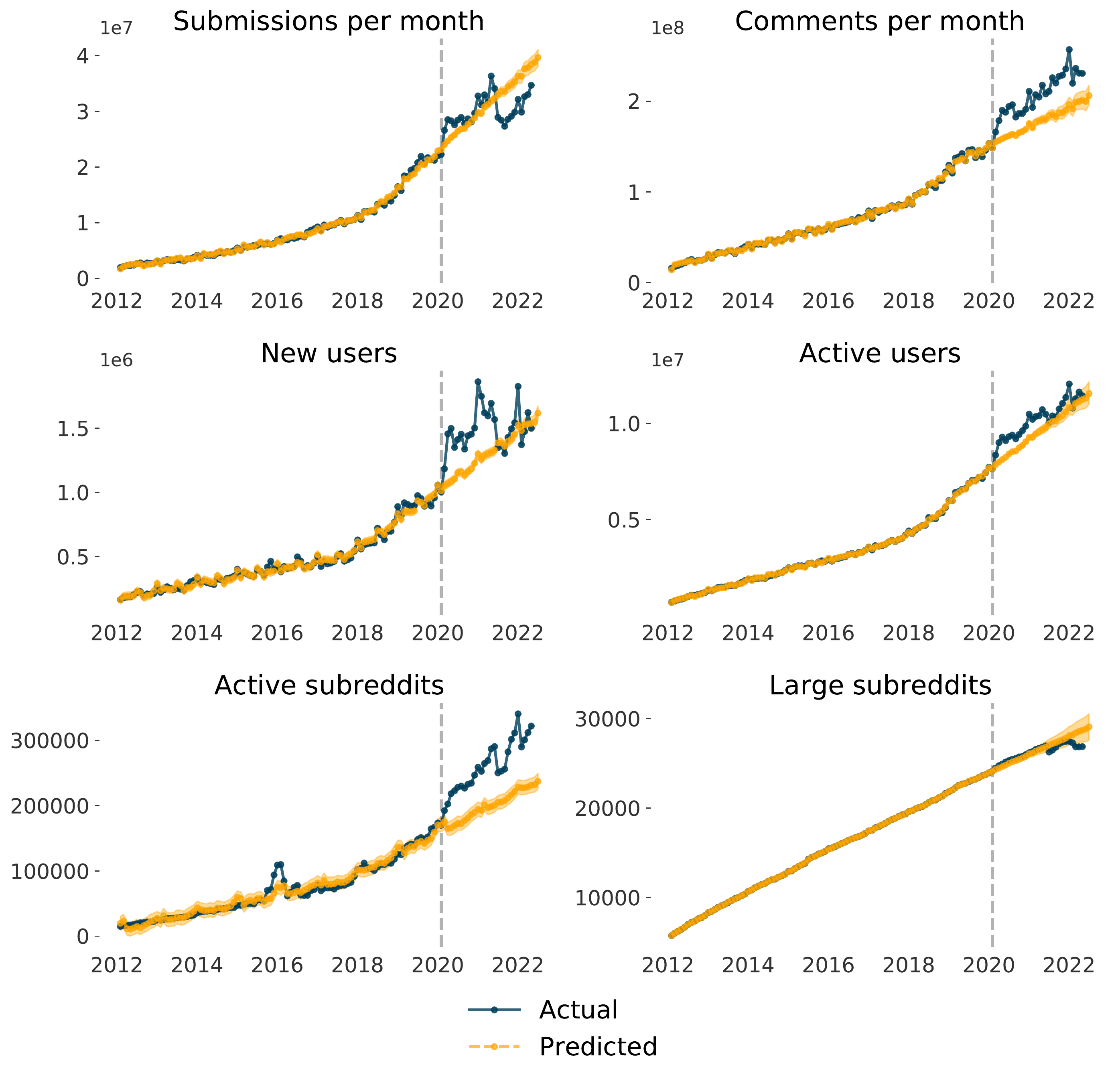}
    \caption{Reddit's growth in a series of metrics. The blue line represents actual growth and the yellow line is a forecast fit with data until 1 February 2020.}
    \label{fig:predicted_features}
\end{figure}


\section{Structural Evolution}\label{structure}
The first series of variables we examine are Reddit's structural components. While this can refer to many platform-level features, we restrict our study to transformations in the user engagement patterns, which include everything from raw activity counts to their habits on the platform. 
This form of meta-analysis is important since it will lay evidence to some of the often-touted claims surrounding the digital transformation which arose as a result of COVID-19. How large was our increase in activity? What drove this spike in activity? And most importantly, what should we expect a post-COVID social media landscape to look like? Moreover, the structural patterns will act as a solid basis for the proceeding content analysis, without which we can make faulty conclusions about the changes we see. 
An example of why this is important is shown in the work of Waller and Anderson where they attribute an observable polarization on Reddit not to a radicalization of existing users, but to a large influx of new users~\shortcite{waller2021quantifying} .

\begin{table}[htbp]
    \centering
    \tiny
    \begin{subtable}{0.23\textwidth}
        \begin{tabular}{clrr}
        \toprule
                & \textbf{Month} &  \textbf{Loss} &  \textbf{z-score} \\
                \midrule
     \multirow{5}{*}{\STAB{\rotatebox[origin=c]{90}{Submissions}}}
        &2020-04-01 &   0.13 &     0.74 \\
        &2019-03-01 &             0.12 &     0.64 \\
        &2016-11-01 &             0.10 &     0.35 \\
        &2020-03-01 &             0.10 &     0.33 \\
        &2019-05-01 &             0.10 &     0.30 \\
        \toprule
     \multirow{5}{*}{\STAB{\rotatebox[origin=c]{90}{New users}}}

            & 2020-04-01 & 0.26 &     1.39 \\
            & 2020-05-01 & 0.25 &     1.31 \\
            & 2018-07-01 & 0.20 &     0.81 \\
            & 2021-01-01 & 0.16 &     0.51 \\
            & 2019-01-01 & 0.16 &     0.50 \\
            
        \toprule
             \multirow{5}{*}{\STAB{\rotatebox[origin=c]{90}{Subreddits}}} 

        & 2016-01-01 &            0.37 &     1.64 \\
        & 2016-02-01 &            0.34 &     1.44 \\
        & 2015-12-01 &            0.32 &     1.32 \\
        & 2015-10-01 &            0.21 &     0.54 \\
        & 2020-05-01 &            0.21 &     0.53 \\
        \bottomrule
        \end{tabular}
    \end{subtable}%
    \begin{subtable}{0.23\textwidth}
        \begin{tabular}{clrr}
        \toprule
                & \textbf{Month} &  \textbf{Loss} &  \textbf{z-score} \\
        \midrule
     \multirow{5}{*}{\STAB{\rotatebox[origin=c]{90}{Comments}}}

        &2020-05-01 &         0.12 &     1.90 \\
        &2020-04-01 &         0.09 &     1.13 \\
        &2020-07-01 &         0.09 &     1.08 \\
        &2020-06-01 &         0.08 & 0.96 \\
        &2018-07-01 & 0.07 & 0.69 \\
        \toprule
         \multirow{5}{*}{\STAB{\rotatebox[origin=c]{90}{Active users}}}

    &2020-04-01 &     0.09 &     1.84 \\
    &2020-05-01 &     0.09 &     1.81 \\
    &2018-07-01 &     0.06 &     0.93 \\
    &2019-03-01 &     0.05 &     0.67 \\
    &2020-03-01 &     0.05 &     0.66 \\
    
    \toprule
    \multirow{5}{*}{\STAB{\rotatebox[origin=c]{90}{Large Subs}}} 
    &2015-07-01 &            0.02 &     1.67 \\
    &2021-11-01 &            0.01 &     0.94 \\
    &2021-10-01 &            0.01 &     0.66 \\
    &2020-04-01 &            0.01 &     0.35 \\
    &2021-12-01 &            0.01 &     0.22 \\
    \bottomrule
    
    \end{tabular}
    \end{subtable}
     \caption{For each metric (submissions, comments, new users, active users, subreddits, and large subreddits) we present the five months where the Prophet model performed the worst (Loss). We additionally show the z-score of the loss as a signal for how wrong the prediction was in relation to all predictions.}
             \label{tab:historic_performance}

\end{table}

\subsection{Macro-level structure}
To gauge how Reddit changed as a whole, we track the evolution of several platform-level metrics. Activity is measured through the number of submissions and comments per month, the user base is measured by the number of new and active users in a given month, and the number of communities is measured by active and large (monthly comments $>$ 500) subreddits. 
To estimate these changes we use the ``effect estimation'' design described in the methods section, where for each metric we fit Prophet on Reddit's historic pre-COVID data and then use this line as an estimate for Reddit's counterfactual growth (no COVID-19). 

The results from this setup are shown in Figure \ref{fig:predicted_features} where we contrast actual (dark blue) \vs predicted (yellow) changes to these metrics. 
We observe large jumps in each of the metrics with the most extreme changes occurring in the number of new users and comments per month. 
In March 2020, the number of submissions jumped by 49\% year-over-year (YoY) from an average growth of 35\% YoY. Comments increased by 27\% YoY, down from an average growth rate of 31\% YoY. This growth in activity was complemented by an influx of new users (52\% YoY) that came to the platform immediately after COVID. 
We similarly found a large increase in the number of subreddits following COVID (47\%), but a small rise in the number of large subreddits. Unlike users, however, this increase had a historical precedent, the period following the 2016 election (125\%). 
A deeper understanding of what leads to a rise in active subreddits warrants further investigation.\footnote{
A possible explanation is that subreddit creation can be used for karma-farming (increasing users' reputation on the platform)} 

To further quantify these changes in relation to Reddit's history, we report the five most difficult months to predict for each metric in Table \ref{tab:historic_performance}. 
Specifically, we trained 89 Prophet models using data from months $t<T$, where $T$ is a variable cutoff parameter encompassing each month between January 2015 and May 2022, and subsequently predicted the outcomes for month $T+1$. 
In other words, instead of stopping the prediction on February 2020 and predicting the future, we now stop the prediction right before each month. This then gives a fair comparison across all-time on the magnitude of different disruptions to the platform. 
We observe that for all metrics, the start of COVID-19 is featured as one of the five largest shocks to the platform; and occupies the top position for the number of comments, active users, and new users. 
This indicates that the changes experienced on Reddit during COVID-19 were truly historic in terms of scale and size. 

Given that COVID-19 was a large momentary shock, we next examine how many of these changes persisted throughout the two-year period following the initial outbreak.
For this analysis, we refer back to Figure \ref{fig:predicted_features} and measure how the difference between the prediction (trained up to Feb. 2020) and the true data for February 2022. 
We argue that if the metrics return to the pre-COVID growth curve then COVID-19's effects were a blip, and if they changed COVID-19 led to a lasting change. 
Interestingly, we find that only the number of comments and active subreddits experienced a lasting magnitude increase.
On the other hand, we observe that the number of submissions experienced a decline in activity (-3.6\%)\footnote{This drop in submissions may be due to some other structural changes on Reddit.} and the user metrics return back to their pre-COVID growth pattern, even after experiencing a large temporary increase. 
This finding acts as a word of caution for the arguments often stated (e.g. by \cite{qureshi2022shifting}) regarding the fundamental digital transformation COVID-19 accelerated. On the contrary, we find that two years on, many of the structural social media metrics returned back to their pre-COVID growth curve.

\begin{figure}[t!]
    \centering
    \includegraphics[width = 3.3in]{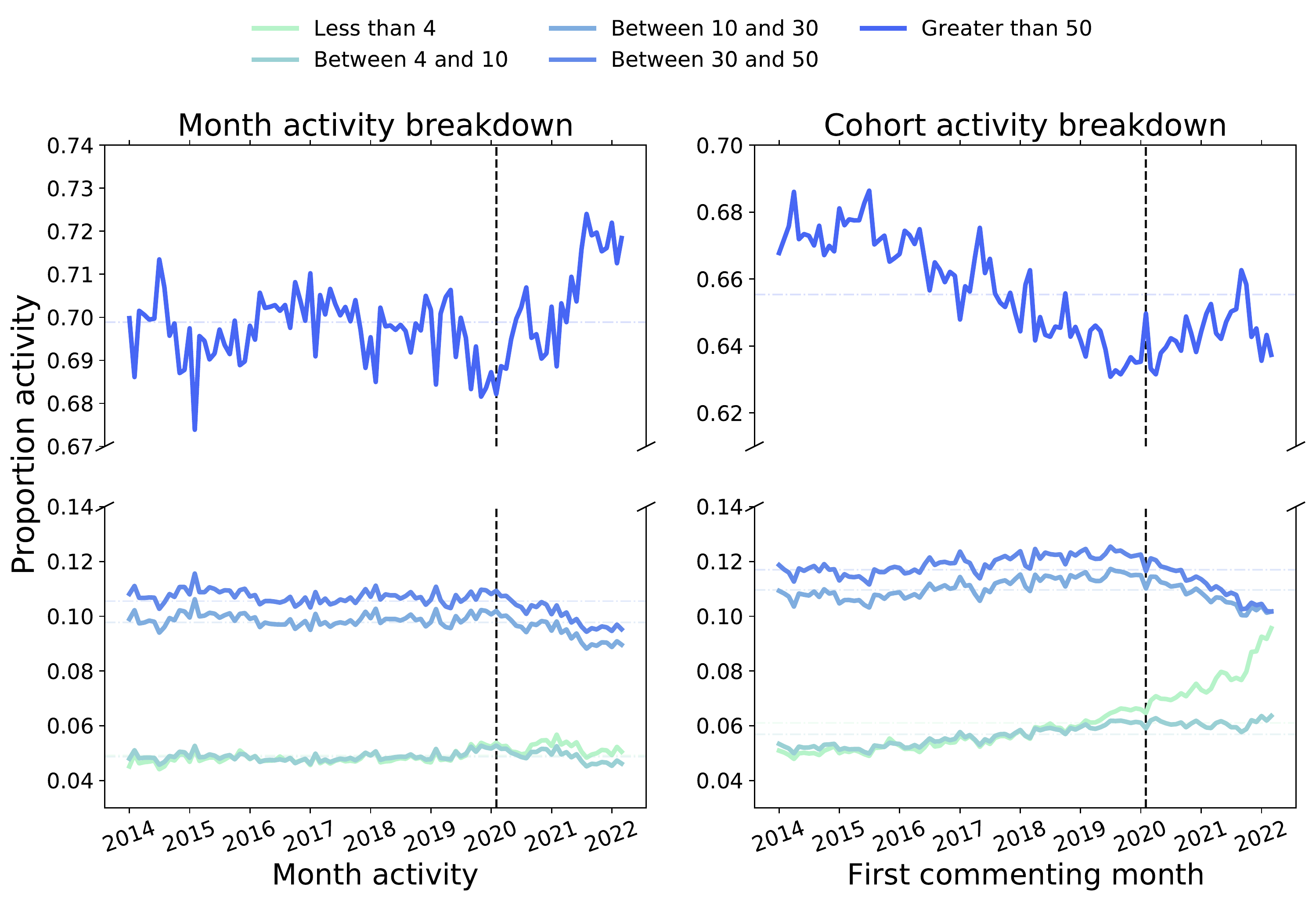}
    \caption{How activity on Reddit changed by different user activity levels and cohorts. The left plot illustrates the proportion of activity in a given month by activity buckets, and the right plot shows the relative activity of users whose first comment were in different months.}
    \label{fig:activty_by_group}
\end{figure}

\begin{figure}[t!]
    \centering
    \includegraphics[width = 2.6in]{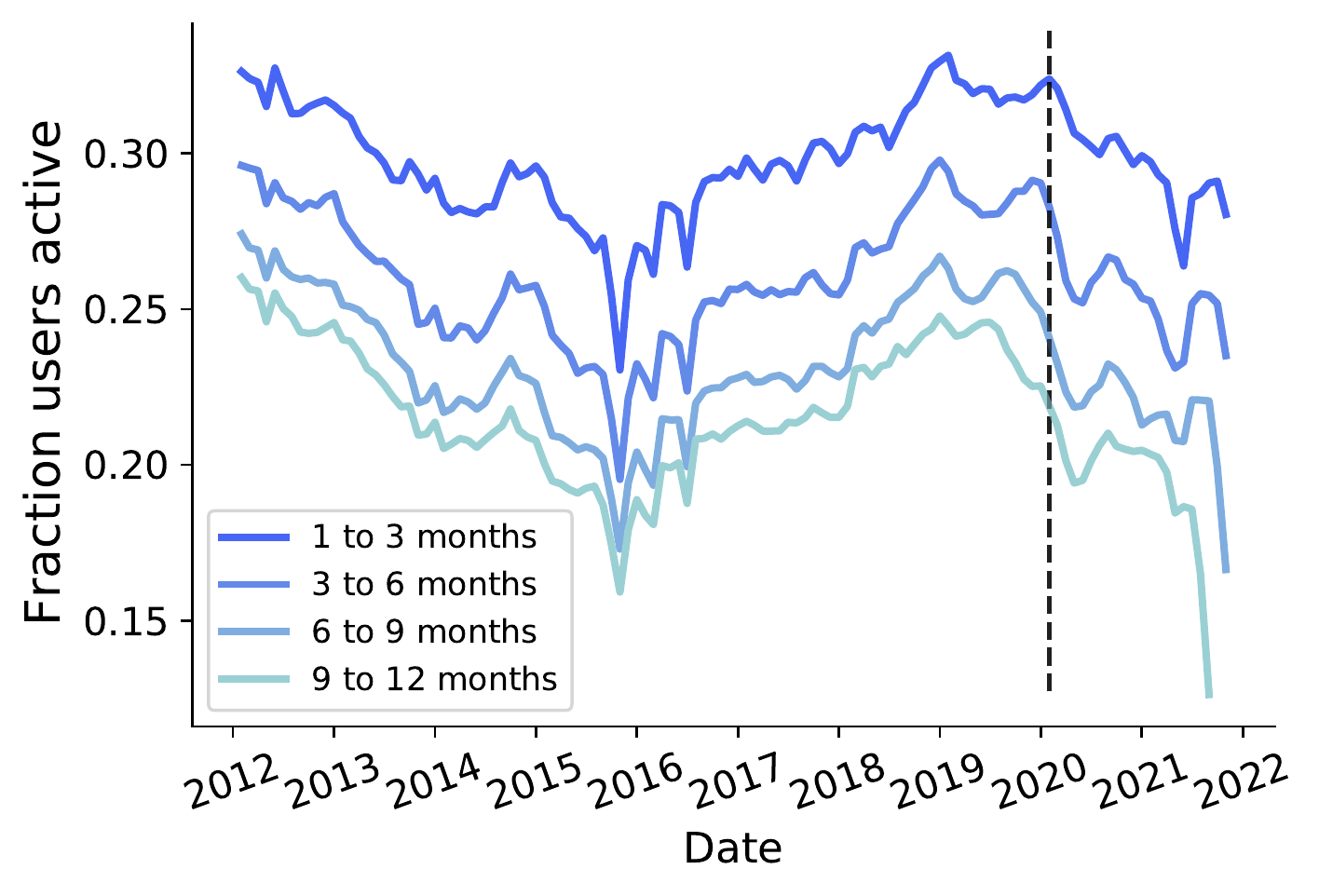}
    \caption{Proportion of users active between the two given months of making their first post.}
    \label{fig:active_between_3_10.pdf}
\end{figure}


\subsection{Micro-level structure}
In the previous subsection, we observed an immediate influx in commenting and posting, alongside a rise in new users, but only a long-run increase in comments and active subreddits. 
While this offers an interesting big-picture understanding of what Reddit went through during COVID, it fails to capture the mechanisms that could have been driving some of those changes. 
In this section, we attempt to uncover what drove two of the changes: (1) commenting behavior exhibited a lasting increase while (2) active users returned to their pre-COVID growth trend.
To understand both of these findings, we turn to the micro-level and examine how different cohorts of users evolved, as well as their propensity to stay active. In the first part, break down users by their activity levels and measure the proportion of comments in a given month by activity bucket and the proportion of a cohort in a given activity bucket. In the second part, we show the fraction of users that remained active for a significant period after their first comment.


\xhdr{Activity buckets} In Figure \ref{fig:activty_by_group} we illustrate how the activity on Reddit changed by different user activity levels. We determine activity level cutoffs by manually examining user commenting distributions and defining intuitive thresholds. More programmatic cutting techniques were attempted, like equal bucket size cuts, but these results tended to be less intuitive and resulted in similar patterns. 
The figure on the left illustrates the proportion of activity in a given month by activity buckets (i.e. ``Greater than 50'' measure the fraction of comments in a given month by users with more than 50 comments that month, and ``Less than 4'' measures the fraction by users with less than 4 comments.)
In general, more activity came from very active users. In fact, users with more than 50 comments represent nearly 73\% activity, a historic high. In addition, there is a relative drop in low to moderate level active users. 
Similarly, the figure on the right plots activity by first-comment month. 
The x-axis represents users' first commenting month, and the y-axis indicates the fraction of activity that came from each activity level within that group. Since each x-value is an aggregate over all months for that cohort, we define their activity level as the mean number of comments during a user's active month. 
We observe that the user cohorts that joined following the COVID-19 outbreak tended to be less active on average than previous cohorts, with more and more of their activity coming from users with less than 4 comments a month on average. 

\xhdr{Activity period} The new users that joined Reddit during the pandemic were also less likely to be active for a longer period of time than users who joined before the pandemic. 
We define a user as active for a longer stretch of time if they comment in windows of $1$ to $3$, $3$ to $6$, $6$ to $9$, $9$ to $12$ months after creating their account. This window technique allows us to equally compare users across different periods of time. Figure \ref{fig:active_between_3_10.pdf} shows how the proportion of active users changes over time. Here we observe a reversal in a previous trend which saw users being active for longer periods of time. Instead, the users that joined during COVID-19 users were less likely to stay active for longer. 
Whereas in the above analysis, we noticed historic changes, this change drop in activity pales in comparison to the drop in new-user activity experienced during the 2016 US presidential elections.

These findings suggest that with the onset of COVID, Reddit experienced a dramatic rise in new users that were not very active, as well as a surge in activity by older cohorts that became more active.
This raises important questions about the needs that Reddit served for different cohorts of users and may suggest that more established users became further enmeshed in the Reddit ecosystem while newer users sought out new forms of entertainment.



\begin{figure}[!t]
        \centering
        \includegraphics[width = 3.2in]{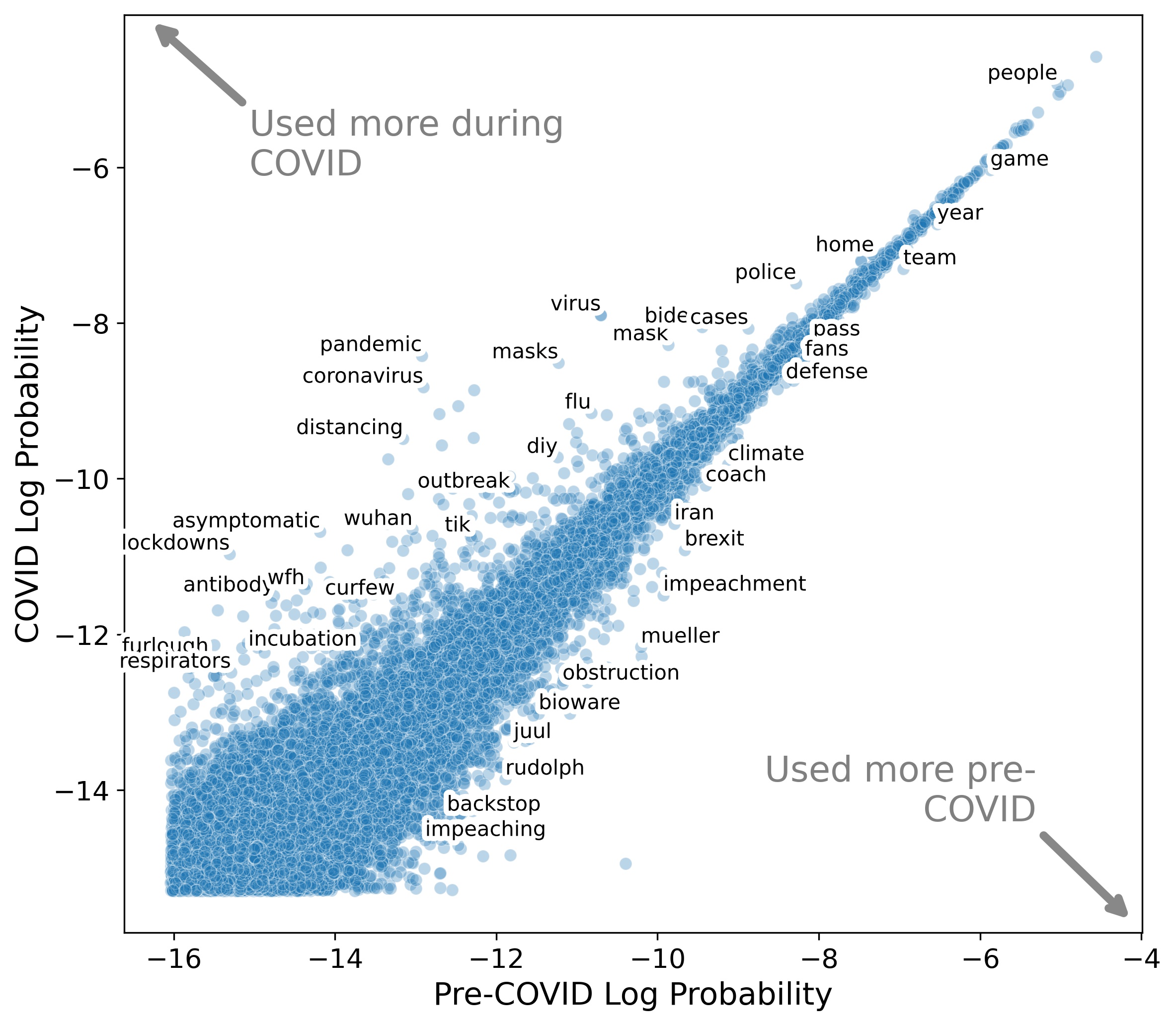}
        \caption{Log odds ratio of language pre- and post-COVID language. The pre-COVID vocabulary is defined as words that occurred twelve months prior to COVID-19 and COVID-19 language is defined as language that emerged in the first five months of the pandemic.}
        \label{fig:log_odds}
\end{figure}%
        ~

\begin{figure}[!t]
    \centering
    \includegraphics[width = 3.3in]{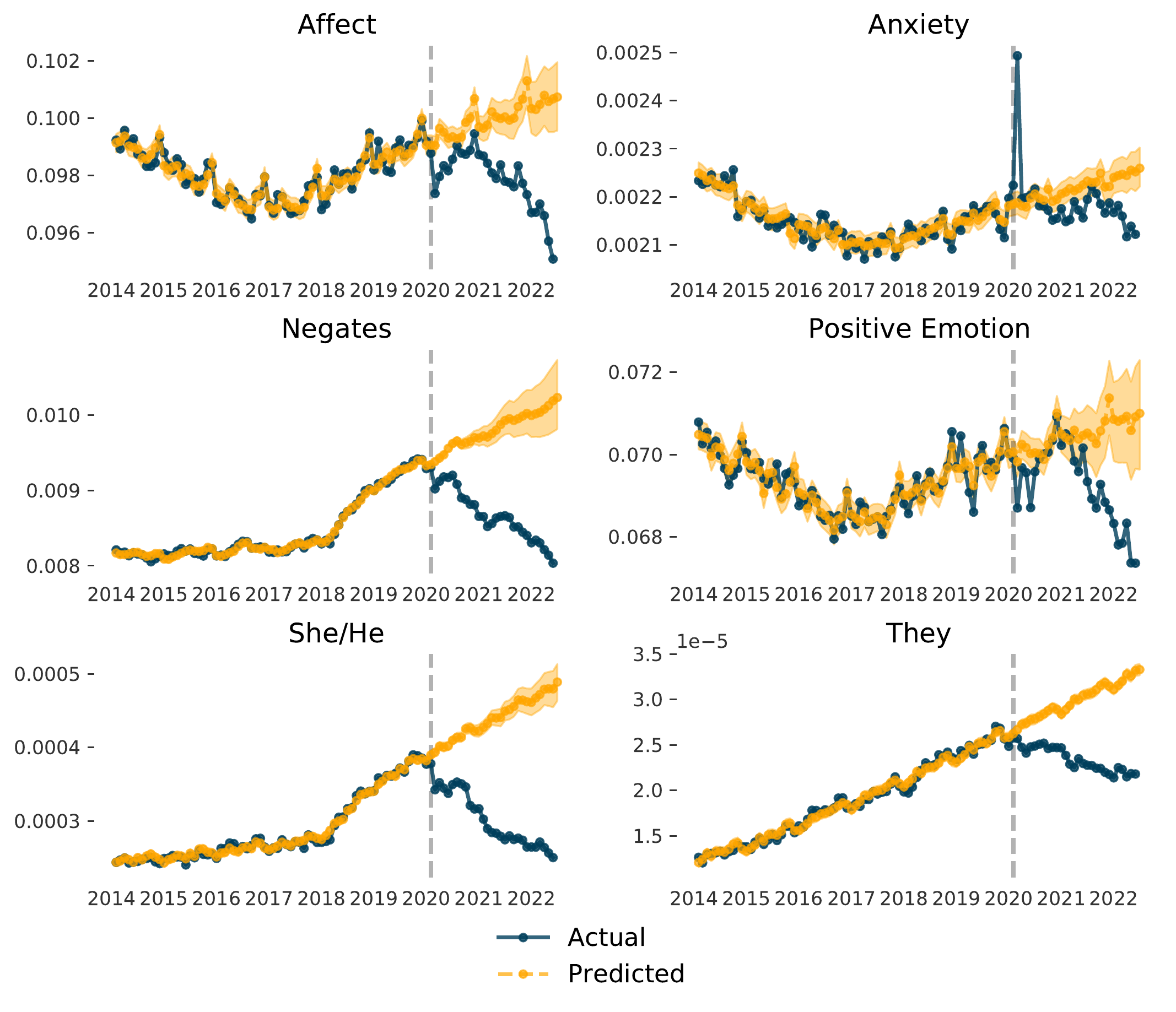}
    \caption{Six LIWC categories that experienced large change throughout COVID-19.}
    \label{fig:liwc}
\end{figure}

\section{Content Evolution}\label{content}

By examining the evolution in user activity we found multiple structural transformations on the platform. 
This analysis, however, failed to capture how the linguistic patterns and topics of interest changed. 
In this section, we supplement the structure with an analysis of content, to gain insight into the role Reddit served as well as a more nuanced understanding of the material it supplied. 
We first examine the macro-level content changes on Reddit, with a particular focus on language and topics. We explore what new language emerged and how it differed from previous language, as well as the types of communities that became more active.
In the second half, we turn to the micro-level by exploring disruptions to the user experience, again through the lens of language (how different cohorts used COVID-related language?) and consumption patterns (did the COVID-wave of users consume quantitatively different content from previous waves of users?).






\subsection{Macro-level content}
We begin by studying how content changed on the macro level; breaking down the analysis into distinct, yet overlapping sections. 
First, we illustrate that language underwent a transformation in terms of the raw occurrences and the fundamental nature of use.
Second, we turn to the subreddit-level changes, by illuminating which topics became popular, and how they differed from previous collections of topics.

\xhdr{Language} 
To study the language we defined a vocabulary of all words with more than 99 uses in any month (an approach used related literature \cite{noble2021semantic}). We then counted the number of times each of the words occurs for each given month between January 2012 and June 2022.

In Figure \ref{fig:log_odds}, we compare the log probabilities of words that saw an increase or decrease in usage before and during the COVID-19 pandemic.
From this plot, we can observe that the new language that emerged was largely COVID-related. For instance, the increases were driven by words directly related to COVID-19, such as ``outbreak,'' ``distancing,'' ``antibody,'' and ``covid,'' as well as indirectly related words like ``diy'' (do-it-yourself).
On the other hand, the words that saw a decrease in usage tended to be related to politics or sports. With the end of the impeachment of American President Donald Trump and the formalization of Brexit in February 2020, we see a decline in the usage of words frequently used during these events, such as ``obstruction,'' ``impeaching,'' ``Mueller,'' and ``Brexit.'' Additionally, with the cancellation of sporting events, there was a relative drop in words like ``pass,'' ``fans,'' ``defense,'' and ``team.'' We also aggregate these words into LIWC categories~\cite{pennebaker2015development} which includes a dictionary of words that capture various psycholinguistic properties. In Figure \ref{fig:liwc} we present six categories that went through dramatic changes throughout COVID-19.\footnote{A figure with all LIWC categories is available in the code repository.} We observe that pronouns like "She/He" and "They", words that "negate" the meaning of a sentence, and affect and positive emotion decline. On the other, there is a sharp temporary spike in anxiety-related language. 

To determine if this change was meaningful, we defined a simple unigram language model over non-COVID-related language. The COVID-language vocabulary was created using existing datasets\footnote{\url{https://www.btb.termiumplus.gc.ca/publications/covid19-eng.html}} and a computational month-by-month comparison of word usage for COVID-related words where we manually annotated the 1,000 words that changed the most from November 2019 to June 2020 and classified them as COVID-related or not.
Then, for each month $t$, we measured the cross-entropy between the probability distribution at month $t$ and $t'$ to evaluate the similarity of the two months in terms of word usage.
We found that in March 2020 there was a significant shift in language with a 0.032 increase in cross-entropy, the largest increase between January 2018 and June 2022. This suggests that the rate of language change accelerated, with certain words becoming more prevalent and others falling out of usage.

\begin{figure}
    \centering
    \includegraphics[width = 3in]{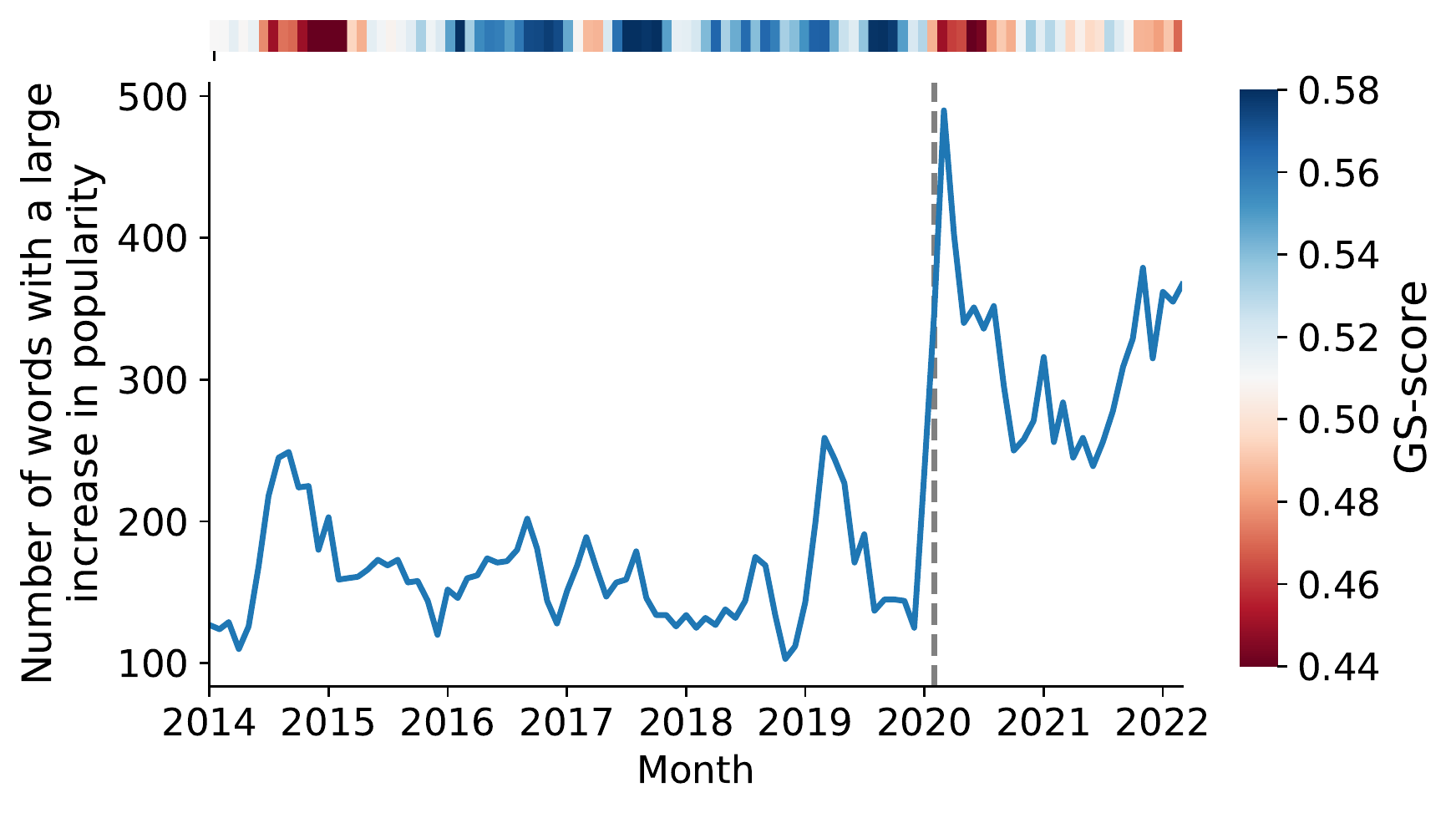}
    \caption{All words that increased in usage from less than the 93rd percentile to above the 94th percentiler. The 1-d heatmap on the top illustrates the GS score of the words that emerged. A lighter indicates a more general vocabulary and a darker color indicates more specialized language.}
    \label{fig:new_words_gs}
\end{figure}
\begin{figure}
    \centering
    \includegraphics[width = 3in]{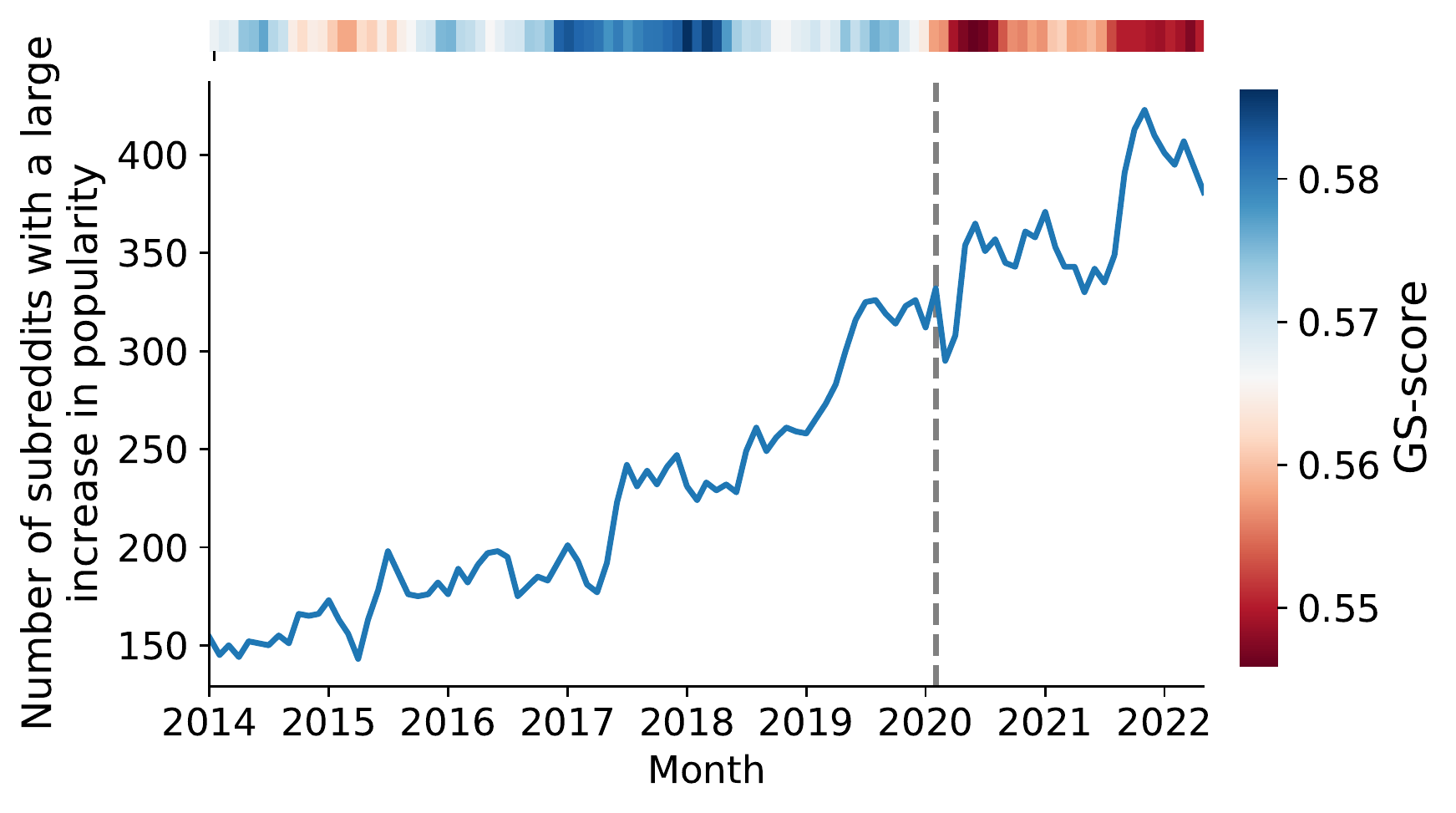}
    \caption{All subreddits that increased in usage by 10 percentile points. The 1-d heatmap on the top illustrates the GS score between the subreddits that emerged. A redder color indicates a more general set of subreddits and a bluer color indicates more specialized subreddits.}
    \label{fig:new_subs_gs}
\end{figure}

One likely cause of this change is the emergence of new words. This is evidenced by a marked increase in the popularity of certain words, as shown in Figure \ref{fig:new_words_gs}, which displays a five-fold increase in the number of popular words. We define a word as popular if it  rises in popularity from below the 93rd percentile to above the 94th percentile, which corresponds to an approximate rise to 1000 times per month, from 500 times or less in February 2020. This trend holds true across different percentile cutoffs. 
Furthermore, the new language that emerged is more semantically diverse than before, as indicated by the 1-d heatmap above the plot. 
We measure semantic similarity within a set of new words by using a static word embedding. 
Explicitly, we map each word in a set of new words onto its GloVe word embedding representation \cite{pennington2014glove} and then calculate the centroid of all new words. We then take the average cosine similarity between all the individual words and the centroid and define this to be the ``generalist-specialist score'' (GS-score) of the new language that emerged~\cite{waller2019generalists}. 
In Figure~\ref{fig:new_subs_gs}, we notice that COVID-19 brought about the most diverse collection of new languages since 2014. 
Language change is often a reflection of broader societal phenomena, and the topics discussed online can provide insight into what people are concerned about. The plot suggests that during the COVID-19 pandemic, users began discussing a wider range of dissimilar concepts. This highlights the significant impact and global reach of COVID-19, as seen in increased discussions of topics such as the economy, healthcare, connection, and well-being.
More interestingly, in other periods where we saw increases in new language, (2016, 2019) the average cosine similarity between the words maintained relatively high, indicating the shocks were less diffuse.

\xhdr{Interest Areas}
We quantify the areas of the platform affected by increased activity identified in the previous structural analysis by creating clusters of subreddits to gain a high-level understanding of Reddit's evolution. We used the clustering method outlined in Waller and Anderson~\shortcite{waller2021quantifying}.
In the work, the authors create and validate a behavioral embedding of Reddit, and then cluster on the subreddit vectors. 
The embedding is created through word2vec applied on subreddits (as words) and users (as contexts), which results in a dense representation of subreddits. 
The proximity between two subreddits indicates a similar set of users is active in both. We then cluster on these dense representations, labeling the clusters based on the communities they represent and their descriptions. A full list of the subreddits that make up these topics is available in our online code repository and a sample of the top five is included in the Appendix.\footnote{\url{https://github.com/CSSLab/reddit-covid-19}}


\begin{figure*}[t!]
    \centering
    \includegraphics[width = 6.4in]{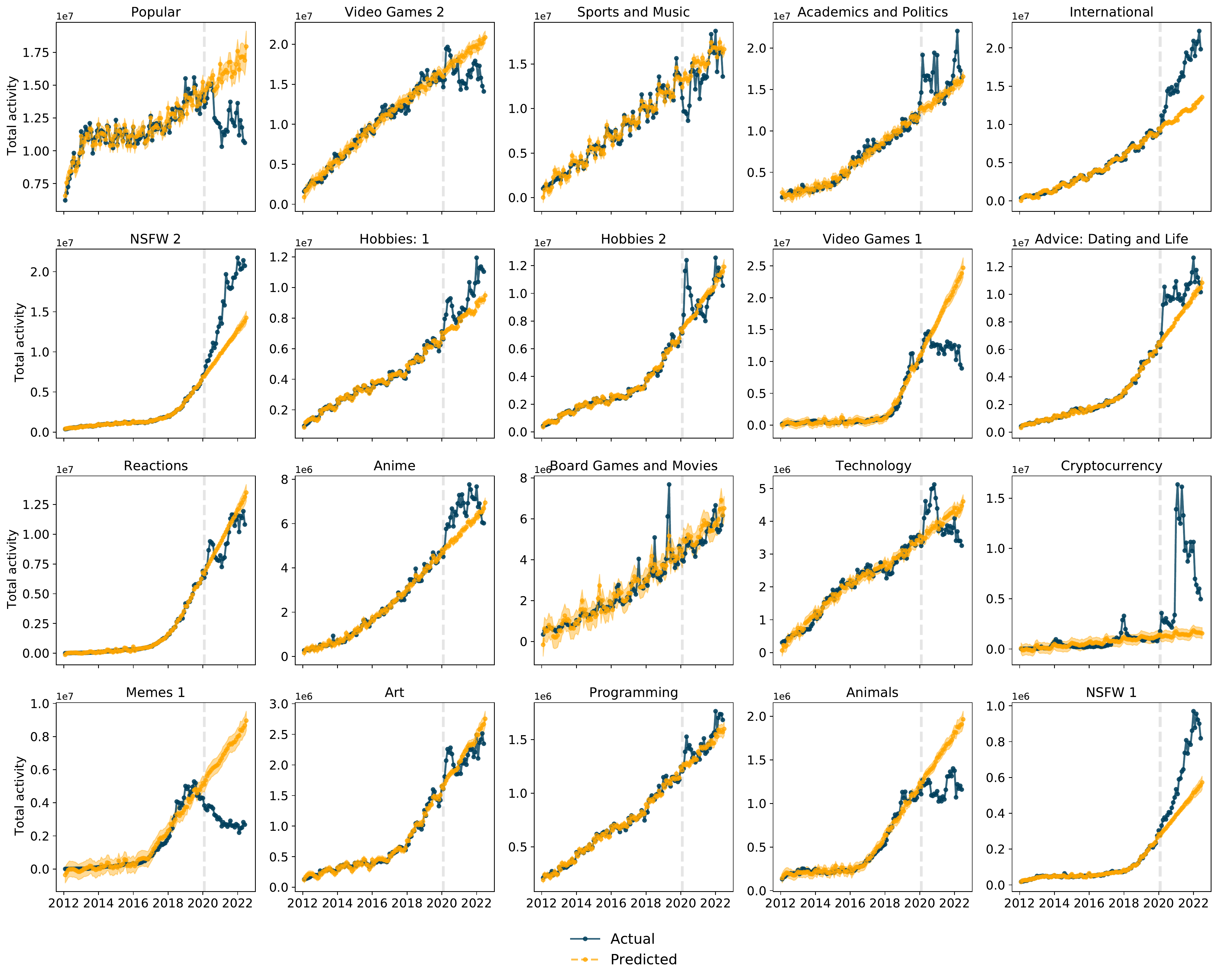}
    \caption{Evolution of topic interests over time. The blue line shows the actual change in activity on a topic whereas the yellow line shows the predicted change. The yellow area surrounding the yellow plot is the 95 confidence interval.}
    \label{fig:topic_activity}
\end{figure*}

Figure \ref{fig:topic_activity} presents activity in these embedding-generated topics with the blue line illustrating actual activity, and the yellow line representing predicted activity. 
Similar to Figure \ref{fig:predicted_features}, the prediction line is trained on the growth trend for each topic until February 2020 and then predicts two years into the future. Consequently, the yellow line represents the expected growth of a subreddit topic prior to COVID. 
We find that there were some clear winners and losers. 
To begin, the International cluster experienced a major growth in activity.
This cluster was composed of a series of global communities that saw a dramatic rise in COVID-19 language and political discussion. These included: r/thenetherlands, r/France, r/India, r/Germany, r/ontario, r/CasualUK, and r/Philippines. As the COVID-19 pandemic continued, this topic saw no sign of dying down and indicates that COVID-19 led to a rapid increase in Reddit's global adoption. 
Another two topics that experienced immediate and sharp increases were (1) Advice: Dating and Life, and (2) Hobbies: 2. These topics were composed of subreddits like r/NoFap, r/datingoverthirty, and r/selfimprovement (1), and  r/getdisciplined, r/selfimprovement, and r/depression (2).  
The biggest winner, however, was NSFW (not safe for work) content. This material in general had little mention of COVID-19-related language but experienced large gains in activity. In fact, during the pandemic NSFW content came to represent almost a two-fold increase in total Reddit activity going up to 13\% from 6.7\% of all comments. 

Interestingly, in addition to some of these major topics that experienced increases, there were also a collection of subreddits that dropped in activity. These subreddits represented traditionally popular content on Reddit and included Animals, Memes, and Minecraft-like video games (Video Games 1). 
Each one of these topics provides a rich lens into how our society evolved throughout COVID-19 and a deep dive into all of these changes is beyond the scope of the paper. However, these initial findings do reveal the complex and ever-encompassing influence the COVID-19 period had on our lives.

We complement Figure \ref{fig:topic_activity} with the subreddit analog of the new words analysis (illustrated in   Figure \ref{fig:new_words_gs}) by looking at the patterns in subreddits that gained a lot of attention. 
To do this we measured the subreddits that increased in activity by 10 percentile points and used the subreddit embeddings to calculate the GS-score (same process as described in the language section). 
We plot this in Figure \ref{fig:new_subs_gs}, and also observe not only a historic rise in the number of subreddits but also that these new subreddits were largely dissimilar from each other. This result is consistent with the change that we saw in language and signals the diffuse impacts COVID-19 had on our lives. Additionally, whereas prior new communities tended to be concentrated around one semantic area, like investing, politics, or sports, the growth that took place during COVID-19 touched on many distinct areas at once.
Contrasting Reddit's macro-changes from a content and structural standpoint, we notice a stark difference. While both have large immediate changes, content changes are far more long-term. 
So while the habits of people on Reddit may not have changed, the content they consumed did change dramatically. 
Consequently, it appears that in the long-run COVID-19 accelerated a change in interest, but not a change in activity.

\begin{figure}[t!]
    \centering
    \includegraphics[width = 3in]{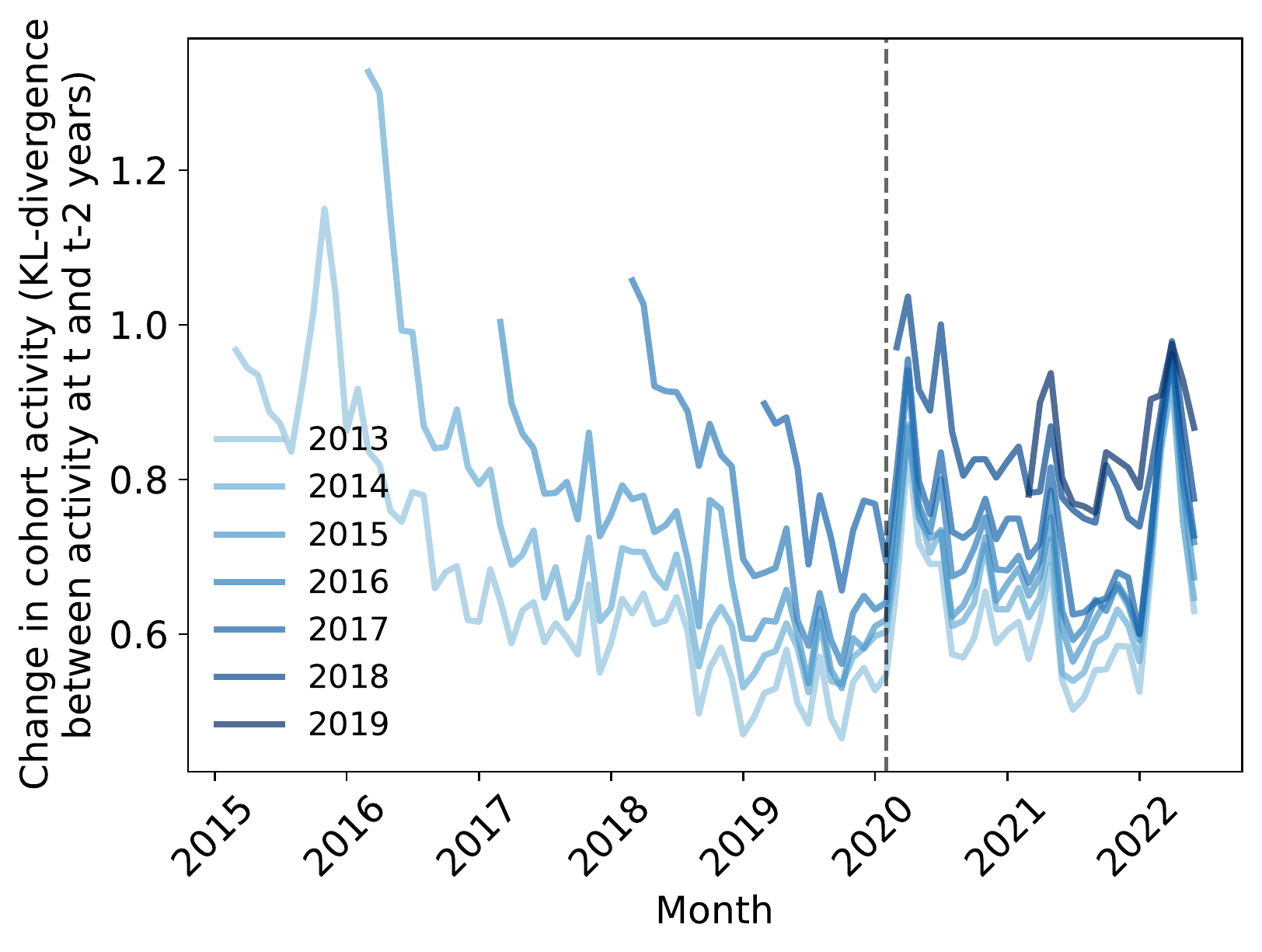}
    \caption{How different cohorts of users were disrupted by COVID.}
    \label{fig:self_kl}
\end{figure}

\begin{figure}[t!]
    \centering
    \includegraphics[width = 3in]{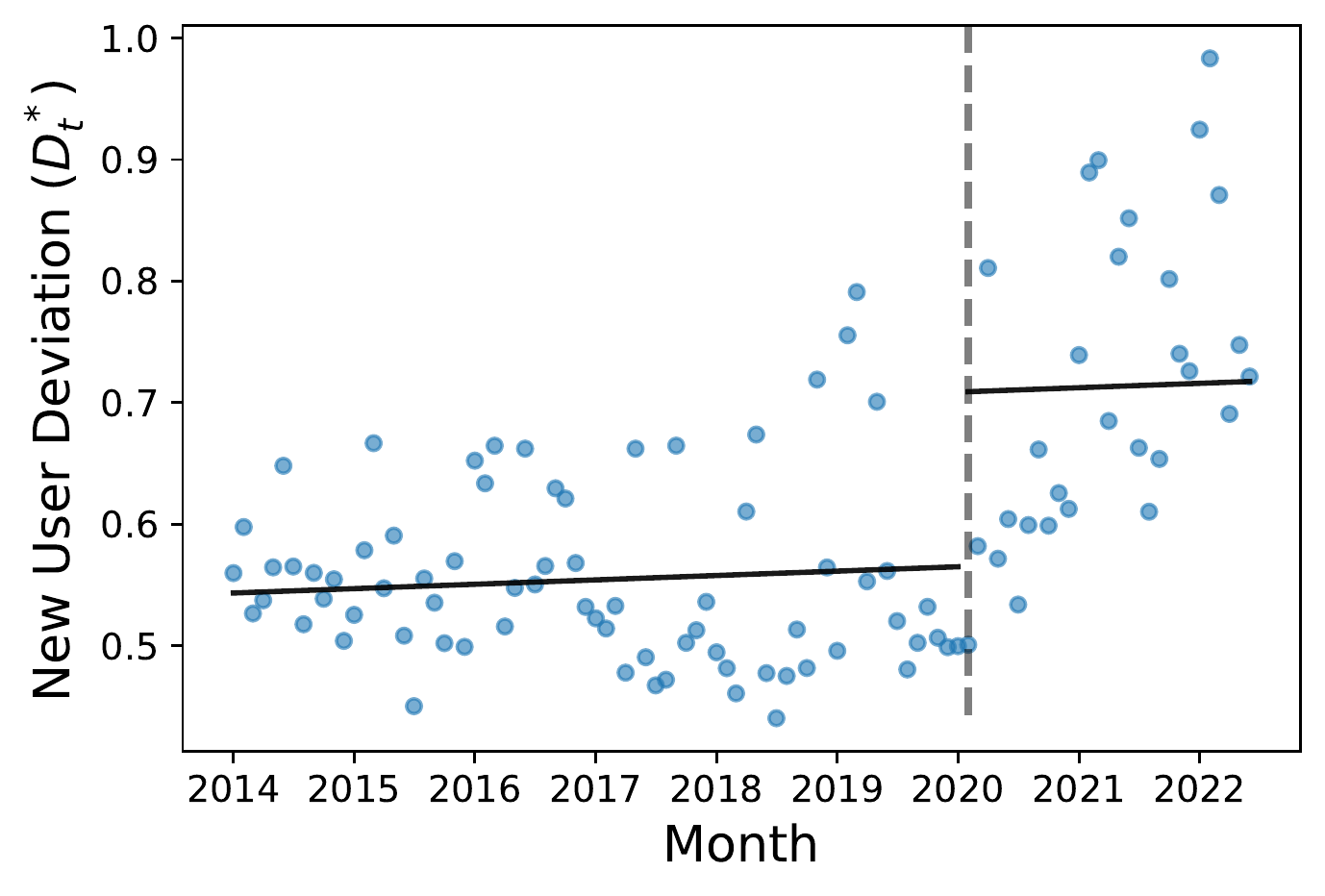}
    \caption{Divergence in user commenting patterns (over subreddits) between users whose account formed in month $t$ and $t-12$ months commenting in month $t$.}
    \label{fig:kl_divergence}
\end{figure}

\subsection{Micro-level content} 
The final section of this paper will complete our proposed 2x2 framework by covering how content changed on the micro-level. Here we connect the various findings we observed above through a study of three questions.
(1) Did all users' activity get disrupted equally? (2) Was the large influx of new users similar to that of new users in the past? (3) To what extent did various cohorts turn to Reddit as a place to discuss COVID? The basic unit of analysis in this section are cohorts. We define a user as being part of cohort $k_m$ if they commented for the first time in month $m$. 

\xhdr{Cohort differences} We begin by examining how different cohorts of users were disrupted by COVID-19.
To measure this, we will compare cohort commenting behavior between time $t$ and $t-12$ months and define a ``deviation,'' $D$, as a disruption from a previously active set of communities. 
Explicitly, for cohort $k_m$ we measure their commenting probability distribution over subreddits for each month $t$, which we will denote $P_{m,t}$. 
Then we will take the KL-divergence for the same cohort across two different periods, and thus the deviation will be $D_{m} = D_{\mathrm{KL}}(P_{m,t} || P_{m,t-12})$
 A high kl-divergence indicates that the communities that the cohort was active in at time $t$ are significantly different from that cohort's activity in time $t-12$; whereas a low kl-divergence suggests relative persistence in interests. 
Figure \ref{fig:self_kl} shows us that in March 2020 there is a historic rise in divergence between all the cohorts; meaning that all users on the platform experienced a shock to their consumption patterns. 
 Older users also returned to their pre-COVID consumption patterns quicker than newer users who typically maintained a higher KL-divergence into the pandemic---indicating that their active communities were more subject to change. 

\xhdr{Differences in new users} In the structure-macro section we observed an unprecedented rise in new users. In this section we explore who these new users were---would they have joined the platform either way or were they a fundamentally new segment of the population? 
Typically when new users join a platform they have a distribution similar to previous cohorts of users. In other words, their activity doesn't deviate from what's considered the norm during a specific time-frame. However, if this deviation begins to grow, then it might indicate that the new users that join the platform exhibit patterns that are divergent from previous cohorts of users, and thus might not represent the ``typical Reddit user.'' 
We operationalize this approach by running a similar setup to the prior analysis only now comparing the distribution in month $t$ between users who were part of cohorts $k_m$ and $k_{m-12}$.
In a similar notation to above, we'll define new user deviation as $D^*_{t} = D_{\mathrm{KL}}(P_{m,t} || P_{m-12,t})$.
We illustrate these divergences in a regression discontinuity in time design in Figure \ref{fig:kl_divergence}. In our context, we consider the treatment at the start of the pandemic with cutoff 1 February 2020 and fit the curve $\hat{y} = \beta_0 + \beta_1 X + \beta_2 t$, where $X$ is an indicator variable equal to $1$ for $t\geq $ ``1 February 2020'' and $0$ otherwise, $t$ is the date and $\hat{y}$ is an estimate of $D^*_{t}$.
The vertical displacement between the two curves estimates the effect of COVID-19 on the new user activity divergence. The jump is equal to $0.1437$ (95\% CI: $[0.08, 0.21]$) and indicates that the mass influx of new users that joined alongside COVID-19 represented a user base whose interests diverged from previous cohorts of users. 

\begin{figure}[t!]
    \centering
    \includegraphics[width= 3in]{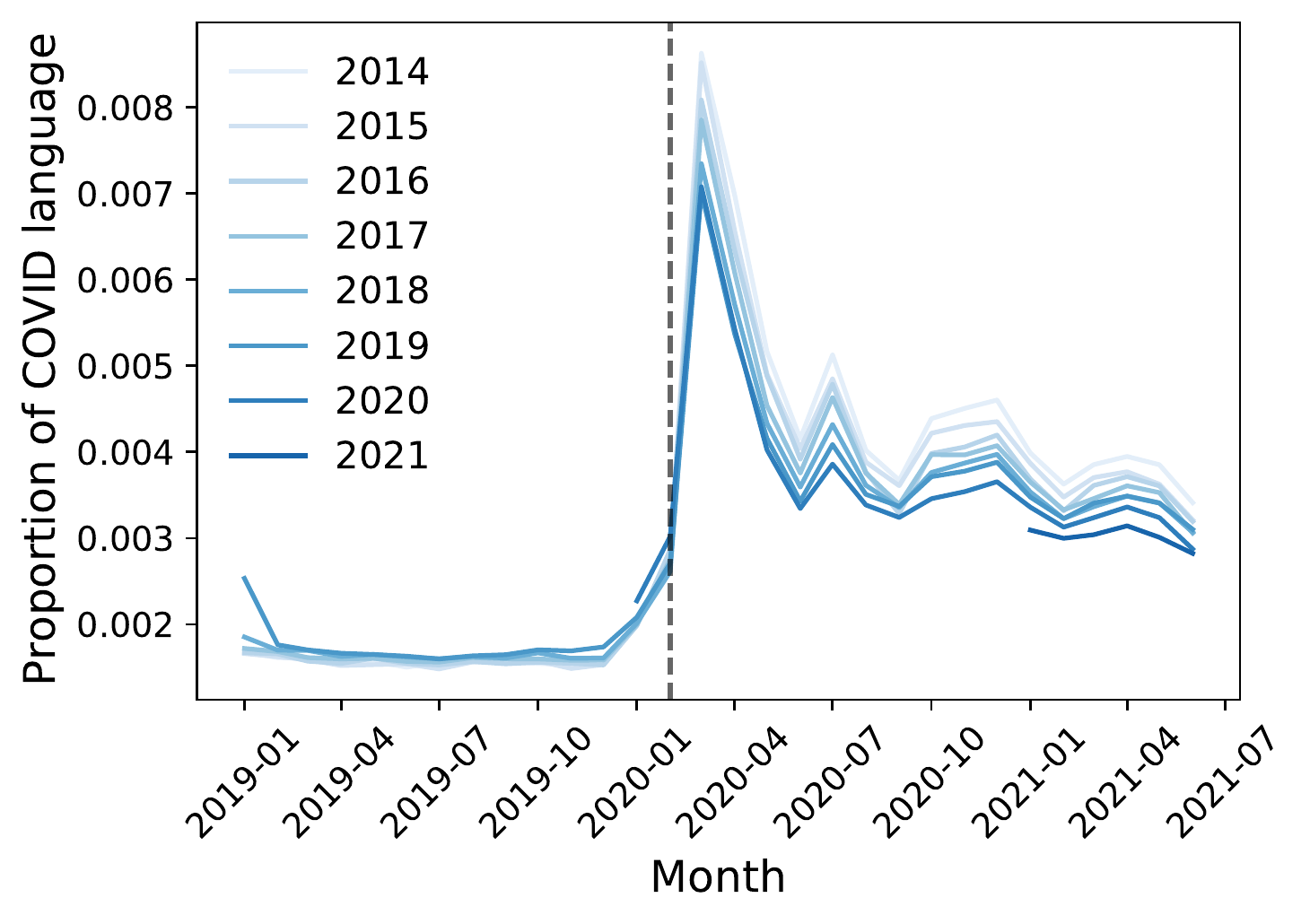}
    \caption{Differences in the fraction of language in a given month that is COVID-19 related across cohorts.}
    \label{fig:cohort_lang}
\end{figure}

\xhdr{COVID related language} 
To better understand the emergence of the new COVID-19 lexicon we ran a cohort-level analysis on a sample of user comments. 
Each user was bucketed into the year they first commented, and then using a 10\% sample of comments from January 2019 to June 2021 we measured the proportion of each cohort's language that was COVID-related. 
We present these proportions by cohort in Figure \ref{fig:cohort_lang}. Immediately, we notice a large jump in COVID-language at the onset of COVID-19. It spiked in March and April 2020 and then saw a leveling off following May 2020. Interestingly, while all cohorts spoke about COVID-19 more, we observe that newer cohorts tended to invoke COVID language less than older cohorts. This may indicate that Reddit served different purposes for well-established users and newer users, and further analysis of this asymmetric pattern may be warranted.


Finally, we'll answer the three questions we prompted this discussion with. First, we find that all cohorts on Reddit were affected by COVID-19, but the older cohorts were able to return to their pre-COVID patterns quicker. Second, the new users that joined Reddit around the time of COVID-19 were fundamentally different from previous sets of new users that joined the platform. As COVID-19 progressed this divergence between new users and users that were one year old continued to increase. Finally, we found a stark difference in how different cohorts started to use COVID-related language. We notice almost a perfect gradient with newer users consistently less likely to use COVID-related language than older cohorts.

\section{Discussion}\label{discussion}

In this paper, we complement previous studies of social media change by presenting a systematic approach to studying platform change and apply it during a focal event. Our framework relies on two axes (structure versus content and macro-level versus micro-level). Structure captures the meta-levels like activity and user composition, whereas content analyzes the actual topics of discussion and language. The macro- versus micro-level change instead examines platform versus user changes. 
This lets us pinpoint whether macro-level changes are driven by new users entering the platform, or whether there are new behaviors emerging across the whole platform. 

\begin{figure}
    \centering
    \includegraphics[width=3.3in]{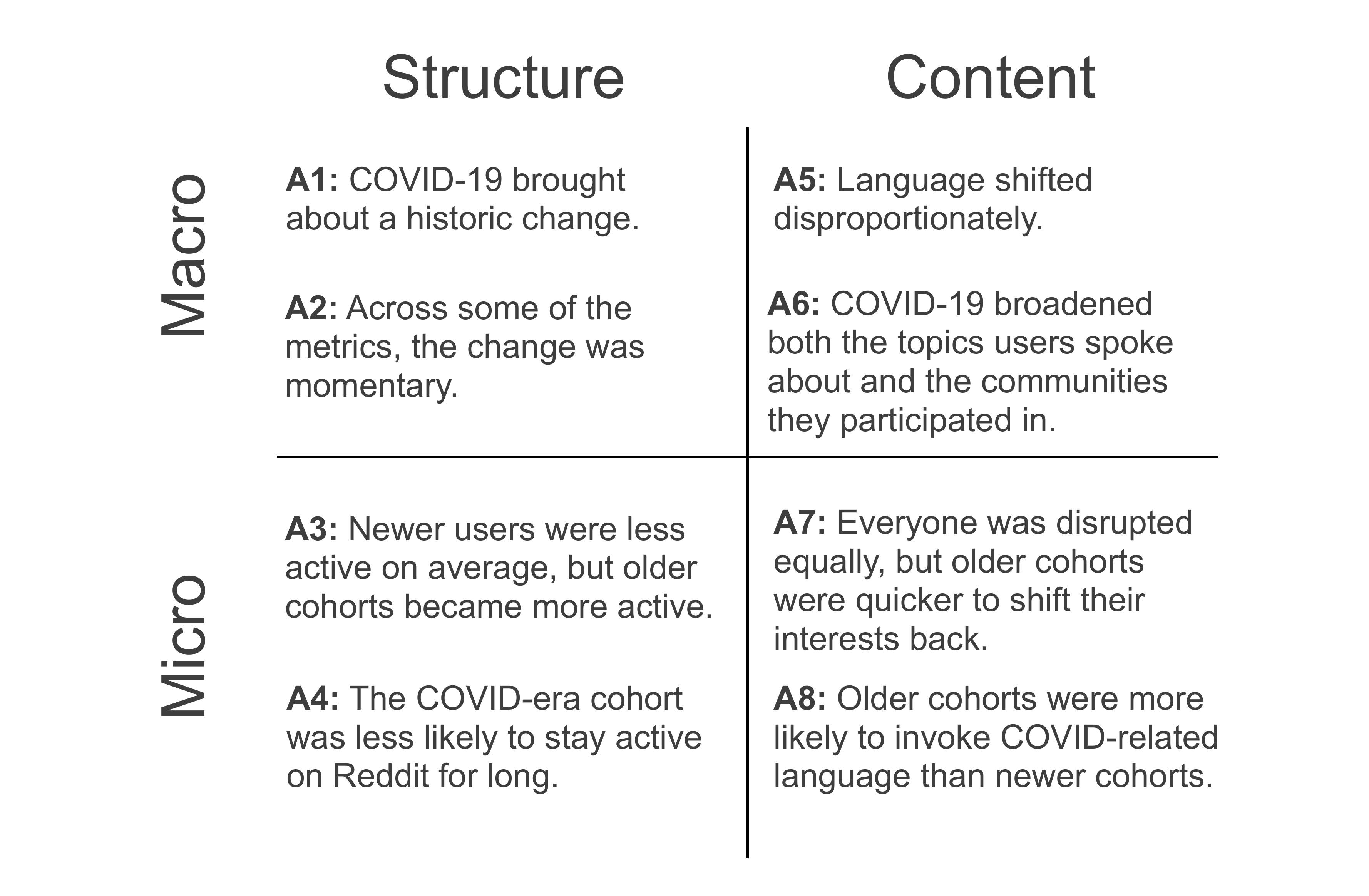}
    \caption{An illustration of the 2x2 (structure \vs content, macro \vs micro) framework as well as the answers to questions we use initially posed.}
    \label{fig:answers}
\end{figure}

At the beginning of this paper, we defined a series of questions that our framework was able to solve. Now in Figure \ref{fig:answers} we summarize the answers to each of these questions. 
In review, Reddit experienced a truly historic increase in activity in relation to its predicted growth. 
We find that across all six structural-macro changes, COVID-19's emergence led to some of the hardest months to forecast. It was one of the five worst months for all metrics covered (comments, submissions, new users, active users, active subreddits, and large subreddits), and the worst month for 4/6 metrics: submissions, comments, new users, and active users. 
In the long run, some of these rises were short-lived. For instance, the number of new users and active users returned to its previous growth curve two years on. But for the number of subreddits and comments per month, Reddit still experiences increases from COVID-19.
On the user level, we notice that these increases were due to older cohorts that became more active; whereas the experience of the new users on the platform was relatively shorter-lived than previous cohorts of new users. 

We then complement the structure analysis by examining how the content of discussions evolved. Content is examined both through the linguistic changes and evolution of areas of discussion (i.e.\ clusters of subreddits). We first notice that following the pandemic, Reddit experiences an unprecedented change in the language that was used on the platform. By defining a language model for each month and predicting language use in other months, we find that the cross-entropy rose after March 2020. This was in part due to a rise in new COVID-related language. This new language that emerged, tended to remain popular for a longer period than the new language that emerged prior to COVID-19 which was usually short-lived and driven by memes and political language. Instead, words like ``coronavirus,'' ``masks,'' ``virus,'' and ``zoom'' all became commonplace and part of our lives. 

We also analyzed content through which people commented by using a custom-made topic classification of subreddits on Reddit. The classification was built through a neural embedding of communities and similarity between communities was defined by similar users commenting in both communities. We found that the topical landscape of which communities became salient also rapidly changed. In general, there was a rapid rise in NSFW, political communities, and investing; and a relative drop in ``traditional'' Reddit content. We connected the linguistic changes with topical changes in activity to track which subreddits experienced a rise in activity by becoming the centers for COVID-19 discussions and which subreddits increased in activity due to people simply staying at home. 
Afterward, aimed to better understand these changes by shifting our attention to the user level to study the topical and linguistic changes. Here we have two key takeaways. First, the older cohorts tended to view Reddit as a place to discuss COVID-19 whereas newer cohorts in general tended to use less COVID-19-related language. Secondly, everyone was disrupted equally by COVID-19, but older cohorts were quicker to shift their interests back. 



There are a few limitations in the current analysis that need to be discussed. To begin, this paper presents a description of the changes that occurred on Reddit during the pandemic. This structure, however, does not allow us to make causal claims of COVID's impacts. Instead, we have rigorously documented what Reddit experienced during the pandemic. For this reason, additional causal structures would have to be added to the analysis to understand the causal mechanisms at play.
With social media occupying an increasingly important role in our lives, it is becoming more important than ever to study and audit these platforms and how they change, since understanding them will help us understand ourselves. 
We believe that the methodology and framework we propose are general enough that they can be applied to different platforms during times of crisis and calm. 
Additionally, one limitation of using a unigram language model is that it misses how the semantics of language change. In general, our approach assumes that over a short period of time, a change in a word is likely driven by an increase in the use of that word and that a significant change in meaning will be accounted for by the use of that word. 

\subsection{Conclusion, Broader Impacts, and Ethical Concerns} This work acts as an empirical foundation for future analyses of social media. We focused on the ``what changed'' question; we believe that this is a necessary precursor to the next set of analyses on ``why'' they changed. In particular, we report a series of interesting, and sometimes surprising, changes that we experienced over the course of COVID-19. Additionally, our approach is general so it can be used in future analyses to study Reddit in times of calm and change. 

There are a couple of possible ethical concerns in our study that we attempt to mitigate. To begin, the Pushift Reddit dump potentially contains deleted comments and posts. But given our aggregated analysis, these harms are minimal. Additionally, no authors were paid or associated with Reddit during the course of the analysis limiting any potential conflicts of interest.

With social media occupying an increasingly important role in our lives, it is becoming more important than ever to study and audit these platforms and how they change, since understanding them will help us understand ourselves. 
We believe that the methodology and framework we propose are general enough that they can be applied to different platforms during times of crisis and calm. 

\xhdr{Acknowledgements} Thank you to Manoel Horta Ribeiro for his helpful comments. As well as NSERC, CFI and ORF for funding the project.

\bibliography{aaai23}

\section{Appendix}
\subsection{Subreddits in Topics}
In Table \ref{tab:cluster_subs} we illustrate the top five subreddits in each cluster.
\begin{table}[t!]
    \centering
    \tiny
    \begin{tabular}{ll}
\toprule
\textbf{Topic name} &                                                                         \textbf{Top 5 subreddits} \\
\midrule
Academics and Politics  &                         politics, conspiracy, worldnews, news,  unpopularopinion \\
\midrule
\multirow{2}{*}{Advice: Dating and Life} &           Tinder, relationship\_advice, RedditSessions, \\ &  distantsocializing, sex \\
\midrule
Animals                 &           Eyebleach, rarepuppers, AnimalsBeingDerps, aww, cats \\
\midrule
Anime                   &           OnePiece, pokemon, anime, Animemes, Genshin\_Impact \\
\midrule
Art                     &           whatisthisthing, fountainpens, Aquariums, gardening, Art \\
\midrule
Board Games and Movies  &           marvelstudios, magicTCG, DnD, StarWars, PrequelMemes \\
\midrule
Cryptocurrency          &           CryptoCurrency, Bitcoin, wallstreetbets, Superstonk, dogecoin \\
\midrule
\multirow{2}{*}{Hobbies 2}              &    AskWomen, relationships, AmItheAsshole, \\ & TwoXChromosomes, Random\_Acts\_Of\_Amazon \\
\midrule
Hobbies: 1              &   personalfinance, Fitness, guns, cars, AskMen \\
\midrule
International           &   de, Cricket, soccer, FIFA, formula1 \\
\midrule
\multirow{2}{*}{Memes 1}                 &   AskOuija, wholesomememes, starterpacks, \\ & insanepeoplefacebook, me\_irl \\
\midrule
NSFW 1                  &   Celebs, selfie, amihot, feetpics, gentlemanboners \\
\midrule
NSFW 2                  &   electronic\_cigarette, TrueFMK, FreeKarma4U, Drugs, gonewild \\
\midrule
Popular                 &   AskReddit, funny, pics, todayilearned, videos \\
\midrule
Programming             &   sysadmin, linux, cscareerquestions, ProgrammerHumor, programming \\
\midrule
Reactions               &   trashy, interestingasfuck, facepalm, PublicFreakout, nextfuckinglevel \\
\midrule
Sports and Music        &   hockey, SquaredCircle, CFB, nba, nfl \\
\midrule
Technology              &   apple, Android, pcmasterrace, MechanicalKeyboards, buildapc \\
\midrule
Video Games 1           &   dankmemes, PewdiepieSubmissions, Minecraft, teenagers, memes \\
\midrule
Video Games 2           &   leagueoflegends, FortNiteBR, DotA2, gaming, DestinyTheGame \\
\bottomrule
\end{tabular}
    \caption{Top 5 subreddits in each cluster.}
    \label{tab:cluster_subs}
\end{table}


\end{document}